\documentclass[conference]{IEEEtran}
\IEEEoverridecommandlockouts
\usepackage{cite}
\usepackage{amsmath,amssymb,amsfonts}
\usepackage{algorithm,algorithmic}
\usepackage{graphicx}
\usepackage{textcomp}
\usepackage{xcolor}
\usepackage[acronym]{glossaries}
\usepackage{blindtext}
\def\BibTeX{{\rm B\kern-.05em{\sc i\kern-.025em b}\kern-.08em
    T\kern-.1667em\lower.7ex\hbox{E}\kern-.125emX}}
\usepackage{booktabs}
\usepackage{multicol}
\usepackage{multirow}
\usepackage{makecell}
\usepackage{tablefootnote}
\usepackage{bm}
\usepackage{import}
\usepackage{physics}
\usepackage[hidelinks]{hyperref}
\usepackage{listings}
\usepackage{pythonhighlight}
\usepackage{nicefrac}

\usepackage{pgfplots}
\pgfplotsset{compat=1.14}
\usepgfplotslibrary{fillbetween}

\usepackage[raggedright]{subfigure}

\usepackage[capitalize,noabbrev]{cleveref}

\definecolor{color_global}{rgb}{0.0, 0.0, 1.0}
\definecolor{color_local}{rgb}{1.0, 0.5, 0.0}
\definecolor{color_classical}{rgb}{0.0, 0.4, 0.0}

\newcommand\copyrighttext{%
  \footnotesize \textcopyright 2024 IEEE. Personal use of this material is permitted.
  Permission from IEEE must be obtained for all other uses, in any current or future
  media, including reprinting/republishing this material for advertising or promotional
  purposes, creating new collective works, for resale or redistribution to servers or
  lists, or reuse of any copyrighted component of this work in other works.}
\newcommand\copyrightnotice{%
\begin{tikzpicture}[remember picture,overlay]
\node[anchor=south,yshift=10pt] at (current page.south) {\fbox{\parbox{\dimexpr\textwidth-\fboxsep-\fboxrule\relax}{\copyrighttext}}};
\end{tikzpicture}%
}

\begin{document}

\newacronym{vqc}{VQC}{variational quantum circuit}
\newacronym{vqa}{VQA}{variational quantum algorithm}
\newacronym{rl}{RL}{reinforcement learning}
\newacronym{qrl}{QRL}{quantum reinforcement learning}
\newacronym{ml}{ML}{machine learning}
\newacronym{qml}{QML}{quantum machine learning}
\newacronym{dnn}{DNN}{deep neural network}
\newacronym{qc}{QC}{quantum computing}
\newacronym{nisq}{NISQ}{noisy intermediate-scale quantum}
\newacronym{pg}{PG}{policy gradient}
\newacronym{qpg}{QPG}{quantum policy gradient}
\newacronym{qnpg}{QNPG}{quantum natural policy gradient}
\newacronym{mdp}{MDP}{Markov Decision Process}
\newacronym{spsa}{SPSA}{simultaneous perturbation stochastic approximations}
\newacronym{fim}{FIM}{Fisher information matrix}
\newacronym{pdf}{PDF}{probability density function}
\newacronym{lse}{LSE}{linear system of equations}

\title{Comprehensive Library of Variational LSE Solvers
    \thanks{
    The research was supported by the Bavarian Ministry of Economic Affairs, Regional Development and Energy with funds from the Hightech Agenda Bayern via the project BayQS.\\
    Correspondence to: nico.meyer@iis.fraunhofer.de
    }
}

\author{
    \IEEEauthorblockN{
        Nico Meyer, Martin R\"ohn, Jakob Murauer, Axel Plinge, Christopher Mutschler, and Daniel D.\ Scherer
    }
    \IEEEauthorblockA{
        Fraunhofer IIS, Fraunhofer Institute for Integrated Circuits IIS, Nürnberg, Germany
    }
}

\maketitle

\begin{abstract}
    Linear systems of equations can be found in various mathematical domains, as well as in the field of machine learning. By employing noisy intermediate-scale quantum devices, variational solvers promise to accelerate finding solutions for large systems. Although there is a wealth of theoretical research on these algorithms, only fragmentary implementations exist. To fill this gap, we have developed the \texttt{variational-lse-solver} framework, which realizes existing approaches in literature, and introduces several enhancements. The user-friendly interface is designed for researchers that work at the abstraction level of identifying and developing end-to-end applications.
\end{abstract}

\copyrightnotice
\glsresetall
\vspace{-0.1cm}  
\section{Introduction.}

Quantum algorithms for matrix arithmetics, in particular solving \gls{lse}, is a strongly studied subject in quantum computing. Most well-known might be the HHL algorithm~\cite{Harrow_2009}, which promises a poly-logarithmic scaling in the system size, compared to the classical polynomial scaling~\cite{Strassen_1969}. However, this improvement is subject to a high sparsity and low condition number of the underlying system. Block encodings can be viewed as a generalization of this concept, with similar promises and requirements~\cite{Low_2019}. The application of these techniques has been studied in various fields, including training neural networks~\cite{Liu_2024} and quantum reinforcement learning~\cite{Cherrat_2023,Meyer_2022a}. However, executing these algorithms requires fault-tolerant and large-scale quantum devices.

With the advent of \glspl{vqa}~\cite{Cerezo_2021}, several variational \gls{lse} solvers have been proposed. The underlying idea is to train a \gls{vqc} to prepare a state that is proportional to the solution of the system. For well-designed and shallow quantum circuit this can potentially be done on \gls{nisq} devices. Our proposed \texttt{variational-lse-solver} library implements several bug fixes and improvements to the approach introduced by Bravo-Prieto et al.~\cite{Bravo_2023}. For systems up to a size of $50$ qubits, they demonstrated a scaling behavior roughly resembling that of HHL. Other studies~\cite{Xu_2021,Patil_2022} have expanded and reformulated this concept. The \texttt{variational-lse-solver} incorporates these approaches or has implementations that are almost identical. The application of the algorithms has been studied in several research efforts~\cite{Trahan_2023,Yoshikura_2023}, yet no implementation was made available.

\section{\label{sec:method}Underlying Concept and Modifications}

The objective is solving a $N \times N$ linear system, comprising $N$ equations and $N$ variables. This can be formulated compactly as finding $\bm{x}$, such that
\begin{align}
    A \bm{x} = \bm{b},
\end{align}
where $A \in \mathbb{C}^{N \times N}$, and $\bm{x},\bm{b} \in \mathbb{C}^{N}$. In practice, there often are additional demands on sparsity and condition number of the system matrix $A$. Bravo-Prieto et al.~\cite{Bravo_2023} propose to variationally prepare a state, that is proportional to the solution $\bm{x}$ of the \gls{lse}. This is formulated in a trainable manner, with $\ket{V(\bm{\alpha})}$ describing the \gls{vqc}-based approximate solution for trainable parameters $\bm{\alpha}$. For the training routine, we need to implement a state $\ket{b}$ that is proportional to $\bm{b}$. Furthermore, a linear combination of unitaries representing $A$ is required, i.e.
\begin{align}
    \label{eq:unitary_decomposition}
    A = \sum_{k=0}^{L-1} c_k A_k,
\end{align}
with all $A_k$ unitary and complex coefficients $c_k$. While decomposing a given matrix $A$ is typically a non-trivial task, there are applications in e.g. quantum chemistry~\cite{Lin_2022} and quantum simulation~\cite{Campbell_2019}, where one has (coherent) access to unitary realizations following \cref{eq:unitary_decomposition}.

\subsection{\label{subsec:modifiedcost}Training with Modified Cost Function}

Training $\ket{V(\bm{\alpha})}$ to be close to the ground truth $\ket{x}$ can be achieved by minimizing the \emph{global} cost function
\begin{align}
    \hat{C}_G = \expval{V(\bm{\alpha}) \left| A^{\dagger} \left( \mathbb{I} - \ket{b}\bra{b} \right) A \right| V(\bm{\alpha})}.
\end{align}
Intuitively, this can also be understood as maximizing the overlap between $A^\dagger \ket{b}$ and $\ket{V(\bm{\alpha})}$, i.e. the ground-truth and the approximated solution of the linear system.
In order to avoid unintentionally minimizing the norm of $\ket{\psi} := A \ket{V(\bm{\alpha})}$, it is common practice to employ a normalized version of the global cost function. Bravo-Prieto et al. suggest the formulation $C_G = \hat{C}_G/\expval{\psi | \psi}$ (see equation 5 in~\cite{Bravo_2023}). We instead propose
\begin{align}
    \label{eq:loss_1}
    C_G &= 1 - \frac{1-\hat{C}_G}{\expval{\psi | \psi}} \\
    \label{eq:loss_2}
        &= 1 - \abs{\expval{b | \Psi}}^2 ~~~~~ \text{with}~ \ket{\Psi} = \frac{\ket{\psi}}{\sqrt{\expval{\psi | \psi}}},
\end{align}
which prevents division of the first summand of \cref{eq:loss_2} by $\expval{\psi | \psi}$ and therefore guarantees convergence of the cost function at a value of $0$. With the same reasoning, we also implement an equivalent re-defined version of the normalized \emph{local} cost function $C_L$ (see equations 6 and 7 in~\cite{Bravo_2023}).

\begin{table*}[htb]
    \centering
    \caption{For the different methods we denote the spatial requirements and the necessary individual circuit evaluations for $n$ data qubits and $m$ terms. Both the raw global and local losses are subsequently composed with the norm following~\cref{eq:loss_1}. The direct method is only suitable for simulation because of its reliance on classical non-unitary matrix manipulation.}\label{tab:framework}%
    \begin{tabular}{@{}l|c||c|c|c|cc@{}}
        \toprule
        ~\textbf{Term} & \textbf{Method} & \textbf{Required Qubits} & \textbf{~~Evaluations}{\footnotesize{$^{\textbf{b}}$}}~~ & \textbf{NISQ Feasibility} & \multicolumn{2}{c}{\textbf{Usage}}~ \\
        \midrule
        \midrule
        ~\multirow{2}{*}{norm $\expval{\psi|\psi}$} & direct & $n$ & $1$ & $-$ & \multicolumn{2}{c}{\textit{default}} \\
        & Hadamard test & $n+1$ & $\nicefrac{1}{2} \left( m^2 -m \right)$ & $\surd$ & \multicolumn{2}{c}{\textit{all non-default}}  \\
        \midrule
        ~\multirow{4}{*}{global $\hat{C}_G$} & direct & $n$ & $1$ & $-$ & \multirow{4}{*}{\texttt{local=False}} & \textit{default} \\
        & Hadamard test{\footnotesize{$^{\textbf{a}}$}} & $n+1$ & $m$ & ($\surd$){\footnotesize{$^{\textbf{c}}$}} & & \texttt{method="hadamard"}~ \\
        & Hadamard-overlap test & $2n+1$ & $\nicefrac{1}{2} \left( m^2+m \right) $ & $\surd$ & & \texttt{method="overlap"~}~ \\
        & coherent{\footnotesize{$^{\textbf{a}}$}} & $n+\lceil \log_2 m \rceil$ & $1$ & $X$ & & \texttt{method="coherent"}~ \\
        \midrule
        ~\multirow{2}{*}{local $\hat{C}_L$} & direct & $n$ & $1$ & $-$ & \multirow{2}{*}{\texttt{local=True~}} & \textit{default} \\
        & Hadamard test & $n+1$ & $\nicefrac{n}{2} \left( m^2+m \right) $ & $\surd$ & & \texttt{method="hadamard"}~ \\
        \bottomrule
    \end{tabular}
    \flushleft
    ~~~~~~~~~~~~\footnotesize{$^{\textbf{a}}$ Implementation with hard-coded components: \url{pennylane.ai/qml/demos/tutorial_vqls/}, \url{pennylane.ai/qml/demos/tutorial_coherent_vqls/}}\\
    ~~~~~~~~~~~~\footnotesize{$^{\textbf{b}}$ Evaluating imaginary terms doubles the number of evaluations for all methods, except direct state evolution.}\\
    ~~~~~~~~~~~~\footnotesize{$^{\textbf{c}}$ Requires implementing multi-control gates, i.e.\ the NISQ-feasibility depends on the decomposition efficiency for a given instance.}
\end{table*}

\subsection{\label{subsec:reducedoverhead}Reduced Evaluation Overhead}

It is possible to evaluate both the norm and the raw (i.e. not normalized) local and global cost functions using either the Hadamard or Hadamard-overlap test~\cite{Bergholm_2022}. We developed two techniques that reduce the number of involved circuits by a factor of up to four:

The computation of the norm $\expval{\psi | \psi}$ makes use of the unitary decomposition of $A$ from~\cref{eq:unitary_decomposition}. With $\beta_{kl} := \expval{\bm{0} | V^{\dagger}(\bm{\alpha}) A_l^{\dagger} A_k V(\bm{\alpha}) |\bm{0}}$ and starting from equation 14 of~\cite{Bravo_2023}, we propose the decomposition
\begin{align}
    \label{eq:norm_1}
    \expval{\psi | \psi} &= \sum_{k}\sum_{l} c_k c^*_l \beta_{kl} \\
                         \label{eq:norm_2}
                         &= \sum_{k} \abs{c_k}^2 + \sum_{k} \sum_{l=k+1} \left( c_k c^*_l \beta_{kl} +  c_l c^*_k \beta_{lk} \right) \\
                         \label{eq:norm_3}
                         &= \sum_{k} \abs{c_k}^2 + 2 \cdot \sum_{k} \sum_{l=k+1} \mathrm{Re} \left( c_k c^*_l \beta_{kl} \right),
\end{align}
where the second step employs $\beta_{kl} = \beta^*_{lk}$. As the terms $\beta_{kl}$ have to be evaluated individually -- using e.g. the Hadamard test for the real part, and a modified version for the imaginary one -- realizing \cref{eq:norm_2} instead of \cref{eq:norm_3} saves half the terms. Additionally, we exploit that the imaginary part of $\beta_{kl}$ can be ignored if $c_k c_l^* \in \mathbb{R}$, leading to a further halving of the required Hadamard tests. An equivalent decomposition exploiting symmetries and real-valued coefficients is also implemented for the raw cost functions $\hat{C}_G$ and $\hat{C}_L$.

\section{\label{sec:experiments}Framework Components and Usage}


The proposed framework is based on \texttt{PennyLane}~\cite{Bergholm_2022} and implements the routines proposed by Bravo-Prieto et al.~\cite{Bravo_2023}, with the modifications discussed in \cref{subsec:modifiedcost}. For training the parameters of the underlying variational circuit, by default an Adam optimizer~\cite{Kingma_2014} with a learning rate of $0.01$ is employed. We found this to be the most stable choice for our experiments, but also ensure customizability for varying setups. Furthermore, the library allows for defining early-stopping criteria, based on the operational meaning of the cost function~\cite{Bravo_2023}.

The following sections cover the details of the implementation, usage, and potential extensions. An overview of the provided functionalities and respective constraints is provided in~\cref{tab:framework}. Additional usage information is also provided in the documentation of \texttt{variational-lse-solver}.

\subsection{Modes of Loading System Matrix}

One integral part of the routine is encoding the system matrix as unitary operations. While there is research effort on the synthesis of circuits given an unitary~\cite{Rietsch_2024}, we approach this topic from a more abstracted view. Our framework allows to provide the decomposition of the system matrix in three different modes: 
\begin{itemize}
    \item A \texttt{circuit} implementing each unitary $A_k$. This allows to integrate custom unitary synthesis techniques.
    \item Explicit \texttt{unitary} representations of $A_k$. This internally employs circuit synthesis tools from \texttt{PennyLane}.
    \item A \texttt{pauli} decomposition $A=\sum_{k} c_k \bigotimes_{i}P_i$, with $P_i \in \left\{ I, X, Y, Z \right\}$. This potentially leads to more terms, but simplifies the internally realized implementation of controlled versions of $A_k$.
\end{itemize} 
Additionally, one can employ direct evaluation of the loss function described in~\cref{subsec:other_evaluation} by only providing the potentially non-unitary \texttt{matrix} $A$. However, this only is feasible in simulation, as it sidesteps the unitary decomposition following~\cref{eq:unitary_decomposition}.

\subsection{Usage Example for Recreating Simple Experiment}\

\begin{figure}[t]
    \usetikzlibrary{calc}
    \centering
    \begin{tikzpicture}
        \begin{axis}[
            name=plot1,
            xlabel=$\textbf{step}$,
            ylabel=$\textbf{loss}$,
            ymode=log,
            xmin=0,xmax=51.5,
            ymin=0.0005,ymax=1.5,
            ytick={1.0, 0.1, 0.01, 0.001},
            yticklabels={,$10^{-1}$,$10^{-2}$,$10^{-3}$},
            label style={font=\footnotesize},
            tick label style={font=\footnotesize},
            ymajorgrids=true,
            axis x line=bottom, axis y line=left,
            width=0.99\linewidth,
            height=3cm,
            legend columns=-1,
            legend style={/tikz/every even column/.append style={column sep=0.1cm, row sep=0.1cm},at={(0.97,1.13)},anchor=east,yshift=-5mm,font=\scriptsize}]
            ]


             \legend{,,,\texttt{global},,,,\texttt{local}}
            
            \addplot+[draw=none,name path=A,no markers] %
            	table[x=step,y=loss_std_upper,col sep=comma]{figures/data/train_global_fun.csv};
            \addplot+[draw=none,name path=B,no markers] %
            	table[x=step,y=loss_std_lower,col sep=comma]{figures/data/train_global_fun.csv};
            \addplot[color_global!40] fill between[of=A and B];
            \addplot[line width=.7pt,solid,color=color_global] %
            	table[x=step,y=loss_avg,col sep=comma]{figures/data/train_global_fun.csv};

            \addplot+[draw=none,name path=A,no markers] %
            	table[x=step,y=loss_std_upper,col sep=comma]{figures/data/train_local.csv};
            \addplot+[draw=none,name path=B,no markers] %
            	table[x=step,y=loss_std_lower,col sep=comma]{figures/data/train_local.csv};
            \addplot[color_local!40] fill between[of=A and B];
            \addplot[line width=.7pt,solid,color=color_local] %
            	table[x=step,y=loss_avg,col sep=comma]{figures/data/train_local.csv};

            \addplot[color=color_classical, domain=0.0:0.001, line width=.7pt] {0.001};
        \end{axis}
            \begin{axis}[
            name=plot2,
            at={($(plot1.south)+(-0.18cm,-1.45cm)$)},
            anchor=north,
            width=0.99\linewidth, height=3cm,
            ymajorgrids,
            label style={font=\footnotesize},
            tick label style={font=\footnotesize},
            xmin=-.5,xmax=7.5,
            ymin=0,ymax=0.25,
            xtick={0, 1, 2, 3, 4, 5, 6, 7},
            xticklabels={000, 001, 010, 011, 100, 101, 110, 111},
            ytick={0.0,0.1,0.2},
            xlabel=$\textbf{basis state}$,
            ylabel=$\textbf{value}$,
            ybar = .05cm,
            bar width = 5.5pt,
            axis x line=bottom, axis y line=left, tick align = outside,
            x axis line style = {-},
            legend columns=-1,
            legend style={/tikz/every even column/.append style={column sep=0.1cm, row sep=0.1cm},at={(0.99,1.34)},anchor=east,yshift=-5mm,font=\scriptsize}]
            ]

            \legend{ground truth,\texttt{global},\texttt{local}}
            
            \addplot[fill=color_classical!90,ybar,no marks,error bars/.cd, y dir=both, y explicit] coordinates {
                (0,0.084)
                (1,0.084)
                (2,0.166)
                (3,0.166)
                (4,0.084)
                (5,0.084)
                (6,0.166)
                (7,0.166)
            };

            \addplot[fill=color_global!80,ybar,no marks,error bars/.cd, y dir=both, y explicit] coordinates {
                (0,0.086) +- (0.007,0.009)
                (1,0.083) +- (0.007,0.011)
                (2,0.165) +- (0.012,0.013)
                (3,0.159) +- (0.011,0.010)
                (4,0.088) +- (0.009,0.008)
                (5,0.085) +- (0.007,0.009)
                (6,0.171) +- (0.009,0.010)
                (7,0.163) +- (0.009,0.009)
            };

            \addplot[fill=color_local!80,ybar,no marks,error bars/.cd, y dir=both, y explicit] coordinates {
                (0,0.083) +- (0.008,0.008)
                (1,0.084) +- (0.007,0.007)
                (2,0.162) +- (0.010,0.009)
                (3,0.166) +- (0.008,0.010)
                (4,0.087) +- (0.009,0.007)
                (5,0.084) +- (0.007,0.007)
                (6,0.166) +- (0.010,0.010)
                (7,0.167) +- (0.010,0.011)
            };

        \end{axis}
    \end{tikzpicture}
    \caption{\label{fig:exp}Results produced with the proposed framework on the \gls{lse} described in~\cref{eq:lse_1,eq:lse_2}. The variational ansatz consists of a depth $d=1$ version of~\cref{fig:circuit}. The training for $50$ steps in the upper plot averaged over $100$ random initializations shows smooth convergence for both loss functions. The evaluation of the final results with $1000$ shots in the lower plot is in good agreement with the normalized ground truth solution. The error bars denote the 25th and 75th percentile over the $100$ trained parameter sets.}
\end{figure}
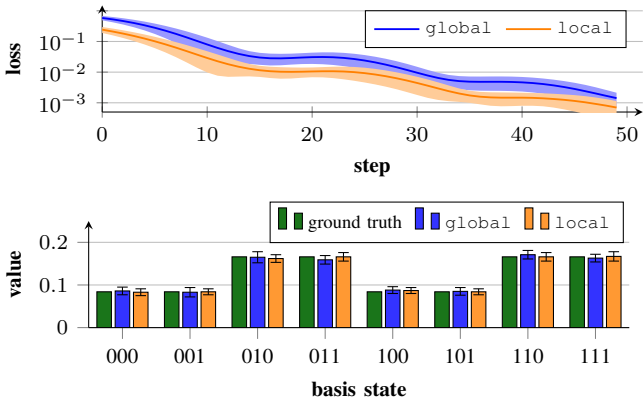

To demonstrate the straightforward usability, we recreate one of the illustrative experiments from Bravo-Prieto et al.~\cite{Bravo_2023}. The task is to variationally solve the linear system defined by
\begin{align}
    \label{eq:lse_1}
    A =& I_0I_1I_2 + 0.2 \cdot X_0 Z_1 I_2 + 0.2 \cdot X_0 I_1 I_2 \\
    \label{eq:lse_2}
    b =& H_0 H_1 H_2 \ket{000},
\end{align}
where we omit the explicit tensor product symbol $\otimes$ for readability. Using the proposed \texttt{variational-lse-solve} framework it is possible to set up and replicate this experiment in just a few lines of code:

\begin{python}
from variational_lse_solver import VarLSESolver

lse = VarLSESolver(["III", "XZI", "XII"],
                   np.ones(8)/np.sqrt(8),
                   coeffs=[1.0, 0.2, 0.2],
                   method="hadamard",
                   local=False,
                   lr=0.01,
                   steps=50)
solution, _ = lse.solve()
\end{python}
Instead of providing the explicit vector representation of $b$ it also would be possible to submit a quantum circuit applying Hadamard gates to the individual qubits. The experimental results depicted in~\cref{fig:exp} demonstrate the effectiveness of the framework. More detailed usage information is provided in the \texttt{README} file.

An example of work integrating the presented library in a more complex routine can be found in Meyer et al.~\cite{Meyer_2024}. More concretely, there the variational algorithm is used to solve linear systems underlying a reinforcement learning setup.

\subsection{Dynamic Circuit Ansatz}

\begin{figure}[t]
    \includegraphics[width=0.99\linewidth]{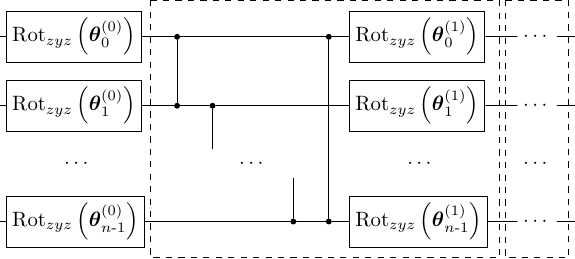}
    \caption{The dynamic ansatz used in the proposed framework, a modified version of~\cite{Patil_2022}. The parameterized gates are realized as $\text{Rot}_{zyz}(\boldsymbol{\alpha}) = R_z(\alpha_2)R_y(\alpha_1)R_z(\alpha_0)$, entanglement is created with nearest-neighbor $CZ$-gates. The initial parameter set $\boldsymbol{\theta}^{(0)}$ is drawn u.a.r. from $\left[ 0, 2\pi\right]$. For increasing the depth, the new parameters are selected as $\boldsymbol{\theta}_i^{(d+1)}=\left( -\alpha, 0, \alpha \right)$, for all qubits $i$, and with $\alpha$ u.a.r. as above. This ensures that the best solution to the \gls{lse} found until that point is not worsened by the modification itself. For depth $d$, the number of trainable parameters therefore is $3n(d+1)$.}\label{fig:circuit}
\end{figure}

In the standard setup the circuit ansatz $\ket{V(\bm{\alpha})}$ is pre-defined and only the parameters $\bm{\alpha}$ are optimized. To allow for a more flexible design, it is possible to use dynamic circuit creation~\cite{Patil_2022}. The proposed framework realizes this strategy using the dynamic ansatz in~\cref{fig:circuit}, with the adjustable parameter of depth $d$ potentially increasing during training. Once the value of the loss function stagnates an additional layer is appended, which parameter initialization realizing a warm start.

This technique is always applied by default, with the possibility to define early stopping criteria, initialization instructions, maximal depth, and more. Additionally, it is also possible to provide custom circuit layouts.

\subsection{\label{subsec:other_evaluation}Additional Evaluation Methods}

We propose and implement additional techniques for the actual evaluation of the loss functions, an overview is provided in~\cref{tab:framework}. First of all, we explicitly realize the Hadamard-overlap test for evaluation of the local cost function, which was briefly mentioned in~\cite{Bravo_2023}. Second, we implement a version of the global cost function that exploits setups with \emph{coherent} access to the unitary realization of $A$~\cite{Kothari_2024,Bergholm_2022}, thus avoiding the need for explicit decomposition and circuit synthesis. Last but not least, we provide a fully differentiable \emph{direct} computation of the losses $C_G$ and $C_L$ following~\cref{eq:loss_1}. This method is not feasible on actual quantum hardware but allows for fast validation and prototyping in simulation.

\glsresetall

\section{\label{sec:conclusion}Concluding Remarks}

We introduce \texttt{variational-lse-solver}, a comprehensive library for research on variational \gls{lse} solvers.
It has to be noted, that it is currently still under debate, in which scenarios, and under which conditions, such techniques are applicable and beneficial. Especially the decomposition of the system matrix, and subsequent realization of the unitary matrices as quantum circuits, requires further attention. To allow for progress toward identifying potentials and further bottlenecks, our framework allows researchers to abstract away those technical details. Due to this, believe this tool will greatly benefit the community and contribute to the progress of quantum software development.

\section*{Code Availability}
The presented python library can be easily installed via \texttt{pip install variational-lse-solver}.
The implementation, with further details on requirements and usage instructions, is also available at
\url{https://github.com/nicomeyer96/variational-lse-solver}
Further information and data is available upon reasonable request.

\bibliographystyle{IEEEtran}
\bibliography{paper}

\end{document}